\documentclass[11pt]{article}
\usepackage[utf8]{inputenc}
\usepackage{geometry}
\usepackage{physics}
\usepackage{dsfont}
\usepackage{amsmath}
\usepackage{amssymb}
\usepackage{graphicx}
\usepackage{babel,blindtext}
\usepackage{xcolor}
\usepackage{hyperref}
\usepackage{cite}
\usepackage{wrapfig}
\usepackage{caption}
\usepackage{subcaption}
\usepackage{bbm}
\usepackage{authblk}
\usepackage{soul}

\title{Thermal quantum correlations and Teleportation in a Graphene Sheet}

\date{}
\author[1]{\small S. Bhuvaneswari}
\author[2]{\small R. Muthuganesan}
\author[1]{\small R. Radha}

\affil[1]{\footnotesize Centre for Nonlinear science(CeNSc),PG and Research Department of Physics, Government College for Women (A), Kumbakonam  Tamil Nadu, India.}
\affil[2]{\footnotesize Department of Physics, Faculty of Nuclear Sciences and Physical Engineering, \\Czech Technical University in Prague, B\u rehov\'a 7, 115 19 Praha 1-Star\'e M\u{e}sto, \\Czech Republic.}

\begin{document}

\maketitle
\section*{Abstract}
The characterization of quantum resources in  dynamical  systems is one of the most important problems to be addressed in quantum information theory. In this article, we investigate the behaviors of quantum correlations and teleportation technique in a  graphene sheet comprising of   disordered electrons in a two-dimensional honeycomb lattice. We use three different measures of quantum correlations  such as entanglement, measurement-induced nonlocality and uncertainty-induced nonlocality.   We study the ground state properties of the graphene sheet from the perspective of quantum correlations. At thermal equilibrium, we show that the band parameter strengthens the quantum correlations whereas  the scattering strength weakens the correlations.  Finally, the impact of the system's parameters on the teleportation technique is also expounded. 

\textit{Keywords}: Entanglement; Quantum correlation; Graphene sheet; Teleportation;

\section{Introduction}
Nonlocality is a fundamental feature of quantum systems which is known to be at the heart of the development of modern quantum technologies. Since the entanglement resources are  crucial for various quantum information processing tasks such as dense coding, teleportation, remote state preparation and key generation \cite{Modi2012}, their characterization and quantification have attracted considerable attention ever since the inception of EPR paradox \cite{Schrodinger1935, Einstein1935} and Bell nonlocality \cite{Bell1, Bell2}. Further, the entanglement also quantifies correlations of nonlocal distributions between spatially separated particles. In the context of quantum information science, the characterization of any quantum system from the perspective of nonlocality is a fundamental and formidable task. The quantification of the nonlocal aspects present in  pure states is more straightforward and entanglement provides the complete spectrum of the nonlocality. On the other hand, there are certain open challenges in the understanding of nonlocality of mixed states and multiqubit systems. In the realm of Bell nonlocality, it is widely accepted that the entanglement captures only a portion of nonlocal character of the system.  To support this, certain tasks were demonstrated in the environment of zero entanglement i.e., other quantum correlations were also detected in some naturally unentangled bipartite systems \cite{Knill1998,Lanyon2008}. To bring out the complete spectrum of nonlocality,  researchers have  identified various measures from  the perspective of the measurements. Nonclassical signatures of quantum systems can be quantified by correlation measures such as quantum discord \cite{Ollivier2001,Dakic2010}, measurement-induced nonlocality (MIN) \cite{Luo2011}, measurement-induced disturbance (MID) \cite{MID2008} and uncertainty-induced nonlocality (UIN) \cite{UIN2014}.  The correlation measures with different notions have unique advantages in the context of information processing task.  In particular, MIN and UIN are considered to be useful  resources for quantum communication \cite{Mattle1996,Li2002} and quantum metrology \cite{Khalid2018,Wu2018,Lu2015} respectively.

The rapid developments which has taken place in the domain of  quantum information processing  has virtually shifted the attention of the  semiconductor industry   towards the fabrication of nanomaterials  which  work in the realm of   quantum regime.   The graphene is one of the most important nanomaterials with a relatively high decoherence time \cite{Fuchs2013}. In view of the implementation of QIP, the behaviors of quantum resources in the graphene system are worthy of   investigation. Graphene is a two-dimensional novel nanomaterial with carbon atoms hexagonally positioned on a honeycomb lattice and is predicted to show interesting magnetic phenomena like intrinsic ferromagnetism \cite{Esquinazi2002,Cervenka2009},   nontrivial electronic properties \cite{Castro Neto2009} etc,. The electronic excitations in graphene arise  due to chiral and massless Dirac fermions and these Dirac Fermions exhibit the properties of quantum electrodynamics leading to the emergence of the anomalous physical phenomenon when subjected to magnetic fields \cite{Gusynin2005,Peres2006} like anomalous quantum Hall effect \cite{Novoselov2005,Zhang2005} and minimal conductivity with charge conductance of $e^2/h$ due to Chiral tunnelling even at the limit of zero temperature \cite{Katsnelson2006,Katsnelson2007}. Entangled resources are studied in solid-state systems for physical realization of quantum computation and communication. Intriguing properties of Graphene have led to the study of its possible applications in quantum memory \cite{Kozikov2012,Wu2012} and quantum teleportation \cite{Asjad2021}. The advantage of using graphene as quantum dots for spin qubits instead of traditional materials like GaAs has also been studied \cite{Recher2010}.  Half-filled energy bands formed due to $\pi$-electrons in graphene lattice are responsible for electron-electron interactions \cite{Miao2007}. These interactions between two electrons can be studied  using "tight binding" approximation \cite{Wallace1947,Suzuura2002}.  Recently, the quantum correlations in graphene system have also been explored from  different perspectives like intrinsic decoherence \cite{Mohamed2022} and thermal effect \cite{Mhamdi2022}.

In this article, we consider the lattice points of graphene sheet as a valid qubit system. We then study the ground state properties of  graphene sheet from the perspective of quantum correlation. We then identify  the correlation measures like entanglement (quantified by the concurrence), trace MIN (T-MIN) and UIN in the ground state. Under  suitable parametric restrictions, it is shown that the system shows  maximal entanglement. At thermal equilibrium, we study the influence of band parameter and scattering strength on the behaviors of  thermal quantum correlations. The effects of system parameters are also studied on the efficacy of the teleportation technique. 

The paper is organized as follows. First, we provide an overview of different measures of quantum correlations in Section. \ref{measures}. We then introduce the model under investigation and its diagonalization in Section. \ref{model}. We then discuss the quantum correlations in the ground state of graphene sheet in Section. \ref{ground}. In Section. \ref{state}, we study the thermal quantum correlations in graphene sheet. The demonstration of teleportation of an unknown state via graphene system is given in Section \ref{Telepor}. Finally, in Section. \ref{cncl}, we conclude with the main results of the paper.


\section{Two-qubit correlation measures}
\label{measures}
In this section, we recall the definition of quantum correlation measures employed here. Let $\rho$ be a bipartite density operator in the separable Hilbert space $\mathcal{H}=\mathcal{H}^A \otimes \mathcal{H}^B$ shared between the marginal states $A$ and $B$, and $\rho^A$ and $\rho^B$ are the density matrices associated with the subsystems in $\mathcal{H}^A$ and  $\mathcal{H}^B$ respectively.

\subsection{Bures entanglement}
 In general, distance quantifier of states is an elegant tool to characterize the quantum correlation contained in a bipartite system \cite{sphener2013}. The Bures distance is a natural candidate to quantify the distance between the states $\rho$ and $\sigma$ \cite{Bengtsson2006} and  is defined as
 
\begin{align}
    d(\rho,\sigma)=2-2\sqrt{\mathcal{F}(\rho,\sigma)} \nonumber
\end{align}

where $\mathcal{F}(\rho,\sigma)=(\text{Tr}[\sqrt{\sqrt{\rho}\sigma\sqrt{\rho}}])^2$ is the fidelity between the states $\rho$ and $\sigma$ and is useful in the quantification of nonclassical correlations such as entanglement and geometric version of discord. The entanglement measure based on the Bures metric is introduced as \cite{Streltsov2010,Marian2003}
\begin{align}
\mathcal{E}(\rho)=2-2\sqrt{\frac{1+\sqrt{1-C(\rho)^2}}{2}}
\end{align}
where $C(\rho)$ represents the concurrence of  the two-qubit state $\rho$. For pure states, the concurrence is defined as  \cite{Hill1997}
\begin{align}
    C(|\Psi\rangle)=|\langle\Psi|\tilde{\Psi}    \rangle|
\end{align}
where $|\tilde{\Psi}    \rangle=\sigma_y^{\otimes2}   |\Psi^* \rangle$ is spin flipped state and (*) denotes the complex conjugation. Similarly, one can define the concurrence for mixed state as 
\begin{equation}
    C(\rho) = \max \{0, \tau_1 - \tau_2 - \tau_3 - \tau_4\},
\end{equation}
where $\tau_i$ are the eigenvalues of matrix $\sqrt{\sqrt{\rho}\tilde{\rho}\sqrt{\rho} }$  arranged in a descending order. Further, the spin flipped matrix is given by $\tilde{\rho}=\sigma_y ^{\otimes2} \rho^* \sigma_y ^{\otimes2}$.  It is worth pointing out at this juncture  that the quantum entanglement $C(\rho)$ varies from 0 to 1 with  the minimal and maximal values corresponding to unentangled (product) and maximally entangled states respectively.  The geometric measure of entanglement is quite crucial  in identifying the optimal decomposition of the quantum state under consideration $\rho$ \cite{Streltsov2010}.
\subsection{Measurement-induced nonlocality}
We have employed measurement-induced nonlocality (MIN) as the second quantifier  of nonlocal correlation. The concept of  MIN
was introduced by Luo and Fu from the perspective of eigenprojective measurements using $ l_2$-norm \cite{Luo2011}. The MIN is defined as the maximal distance between the quantum states $\rho$ under consideration and the corresponding state after performing a local eigenprojective measurement on one of the subsystems, say $A$ i.e.,\cite{Luo2011}
\begin{equation}
	N_{p}(\rho)=~^{\text{max}}_{\Pi ^{A}}\| \rho - \Pi ^{A}(\rho )\|_p ^{p},
\end{equation}
where $\|\mathcal{O}\|_p $ is $l_p$-norm of operator $\mathcal{O}$ and the maximum is taken over the locally invariant projective measurements on subsystem $A$ which does not change the marginal state $\rho^A$.  The post-measurement state is defined as $\Pi^{A}(\rho) = \sum _{k} (\Pi ^{A}_{k} \otimes   \mathds{1} ^{B}) \rho (\Pi ^{A}_{k} \otimes    \mathds{1}^{B} )$, with $\Pi ^{A}= \{\Pi ^{A}_{k}\}= \{|k\rangle \langle k|\}$ being the projective measurements on the subsystem $A$. For $p=2$, the above measure is Hilbert-Schmidt norm based MIN. On the other hand, if $p=1$, the measure is trace distance based MIN.   Eventhough the measures based on $l_2$-norm are easy to compute   and can be experimentally  realized  \cite{Jin2012,Passante2012,Girolami2012}, they  do not represent the  faithful measure of quantum correlation \cite{Piani2012}. Hence, we study the trace MIN in the physical system under consideration. The general expression of a two-qubit density operator acting in the separable composite Hilbert space
$\mathcal{H}^A\otimes\mathcal{H}^B$ is 
\begin{equation}
	\rho=\frac{1}{4}\left(\mathds{1}^{A} \otimes \mathds{1}^{B}+ \mathbf{s} \cdot  \boldsymbol\sigma \otimes \mathds{1}^{B} + \mathds{1}^{A} \otimes \mathbf{r}\cdot\boldsymbol\sigma+\sum_{m,n=1}^{3} t_{mn} \sigma_{m} \otimes \sigma_n\right), \label{Equations} 
\end{equation}
where $\sigma_j ~(j=1,2,3)$ are the Pauli operators. The vectors $\mathbf{s}$ and $\mathbf{r}$ are real with their components being $s_j=\text{Tr}(\rho (\sigma_j\otimes \mathds{1}^B))$,   $r_{j}=\text{Tr}(\rho (\mathds{1}^A\otimes \sigma_{j} ))$ and $t_{mn}=\text{Tr}(\rho(\sigma_{m} \otimes \sigma_{n}))$ being the matrix elements. Without loss of generality, the canonical form of the Fano parametrization of the density operator (\ref{Equations})  is written as 
\begin{equation}
	\rho=\frac{1}{4}\left(\mathds{1}^{A} \otimes \mathds{1}^{B}+ \mathbf{s} \cdot  \boldsymbol\sigma \otimes \mathds{1}^{B} + \mathds{1}^{B} \otimes \mathbf{r}\cdot\boldsymbol\sigma+\sum_{j= 1}^{3} c_{j} \sigma_{j} \otimes \sigma_{j} \right) \label{state2} 
\end{equation}
with $c_j=\text{Tr}(\rho(\sigma_{j} \otimes \sigma_{j}))$. With the above preparation, the closed formula of trace MIN $N_{1}\mathcal{(\rho)}$ is given as \cite{tracemin}
\begin{equation}
	N_{1}\mathcal{(\rho)}=
	\begin{cases}
		\frac{\sqrt{\chi_{+}}~+~\sqrt{\chi_{-}}}{2 \Vert \textbf{x} \Vert_{1}} & 
		\text{if} \quad \textbf{x}\neq 0,\\
		\text{max} \lbrace \vert c_{1}\vert,\vert c_{2}\vert,\vert c_{3}\vert\rbrace &  \text{if} \quad \textbf{x}=0,
	\end{cases}
	\label{TMIN}
\end{equation}
where $\chi_\pm~=~ \alpha \pm 2 \sqrt{\tilde{\beta}}~\Vert \textbf{x} \Vert_1 ,\alpha =\Vert \textbf{c} \Vert^2_1 ~\Vert \textbf{x} \Vert^2_1-\sum_i c^2_i x^2_i,\tilde{\beta}=\sum_{\langle ijk \rangle} x^2_ic^2_jc^2_k,~ \vert c_i \vert $ are the absolute values of $c_i$ and the summation runs over cyclic permutation of 
$\lbrace 1,2,3 \rbrace$.

\subsection{Uncertainty-induced nonlocality}
Next, we employ another important skew information-based measure, namely  uncertainty-induced nonlocality (UIN) and it can be considered as an updated version of MIN \cite{Luo2011}.  This quantum correlation measure is defined as the maximal skew information achievable between a given bipartite quantum state $\rho$ and a locally commuting observable $R^a\otimes\mathds{1}^b$. It is defined as \cite{UIN2014}
\begin{align}
\mathcal{Q}(\rho)=~^{\text{max}}_{R^{A}} ~~\mathcal{I}(\rho, R^A\otimes\mathds{1}^B).
\end{align}
where the maximum is taken over the set of all observables and $\mathcal{I}(\rho, R^A \otimes \mathds{1}^b)=-\frac{1}{2}\text{Tr}([\sqrt{\rho}, R])$ represents the  skew information of the state $\rho$ with respect to the observable $R^A$. UIN also satisfies all the necessary axioms of a valid measure of bipartite quantum correlation and is reduced to entanglement monotone for $2\times n$ dimensional pure states. It also has a closed formula for a qubit-qudit system \cite{UIN2014} as 
\begin{equation}
	\mathcal{Q}(\rho)=
	\begin{cases}
	1-\frac{1}{\| \mathbf{s'}\|^2} \mathbf{s'}\mathcal{W}\mathbf{s'}^t & 
		\text{if} \quad  \mathbf{s'} \neq 0,\\
		1-\omega_{\text{min}} &  \text{if} \quad \mathbf{s'}=0,
	\end{cases}
	\label{TMIN}
\end{equation}
where $\omega_{\text{min}}$ is the lowest eigenvalue of the  matrix $\mathcal{W}$ with the matrix elements $\omega_{ij}=\text{Tr}[\sqrt{\rho}(\sigma_i\otimes \mathds{1})\sqrt{\rho}(\sigma_j\otimes \mathds{1})]$ and  the components of $\mathbf{s'}$ are $s'_i=\text{Tr}[\sqrt{\rho}(\sigma_i\otimes \mathds{1}^B)]$. 
\section{Graphene Sheet}
\label{model}
\begin{figure*}[!ht]
\centering\includegraphics[width=0.85\linewidth]{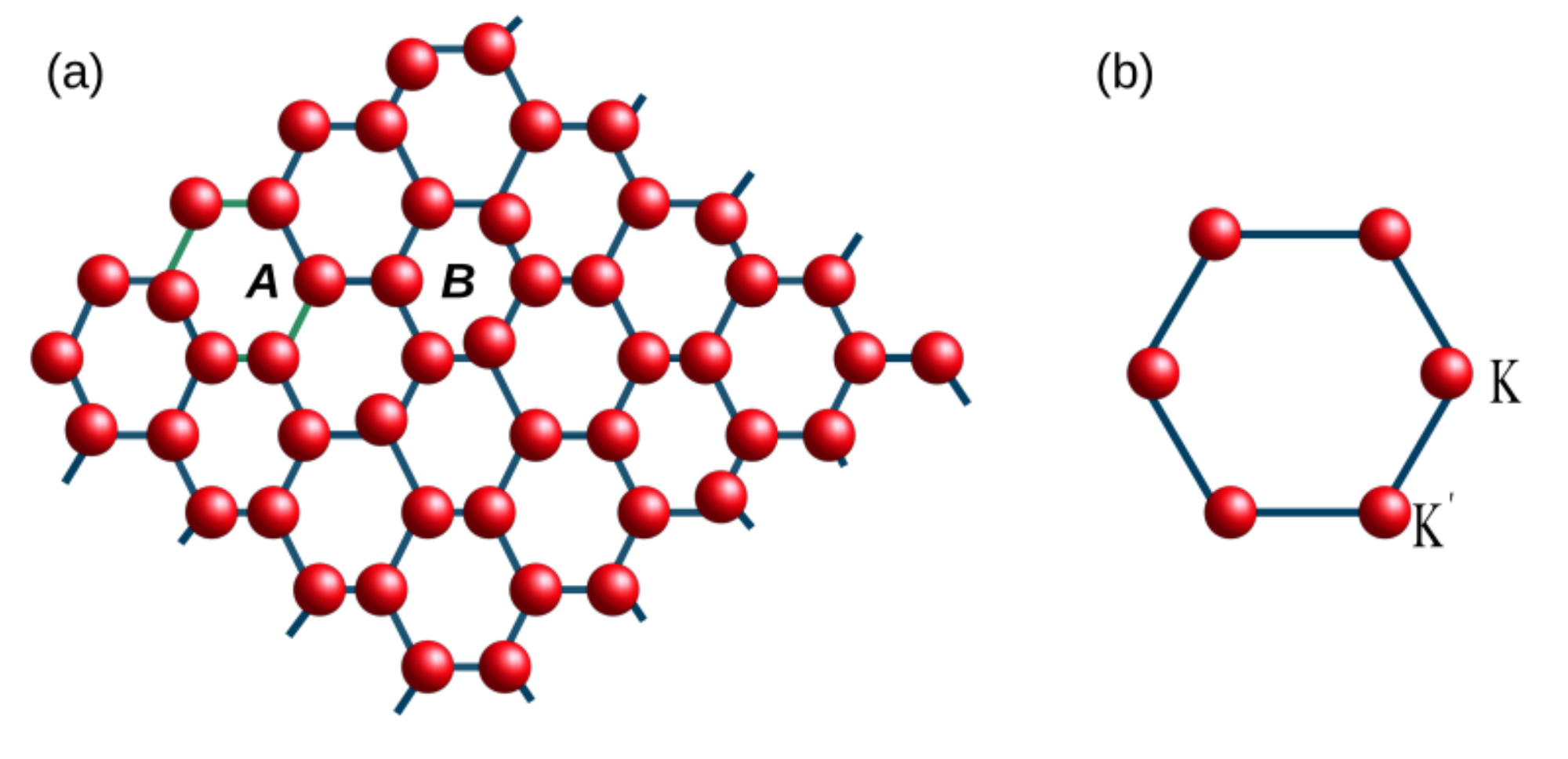}
\caption{(a) A schematic representation of the Honeycomb lattice of graphene sheet with carbon atoms. (b) Dirac points $K$ and $K'$ on a cell.}
\label{graphene}
\end{figure*}

Here, we consider the disordered electrons of a graphene sheet in a two-dimensional honeycomb lattice serving as the physical carrier of quantum information. The graphene sheet consists of multiple primitive cells each with two sublattices denoted by A and B as shown in Fig. \ref{graphene}(a) \cite{Miao2007,Wallace1947, Suzuura2002}. Further, sublattices of the two sites are represented by up $|\uparrow\rangle$ and down $|\downarrow \rangle$  pseudo-spins $\vec{\sigma}_{\alpha}(\alpha=x,y,z)$ with $\sigma$ being  the Pauli matrices. These two sites correspond to two inequivalent valleys occurring at  two different Dirac points $K$ and $K'$  represented by another spin vector $ \vec{\kappa}=(\kappa_x~ \kappa_y~\kappa_z) $ which is described by the states up $| 1\rangle$ and down $|0 \rangle$ as shown in  Fig. \ref{graphene}(b). 

Within the effective-mass approximation, the Hamiltonian around the Dirac points is given by \cite{Suzuura2002,McCann2006}
\begin{equation}
     \hat{H} = \eta \left[\hat{\eta}_x\left(\sigma_x \otimes\mathbbm{1}^B\right)+\hat{\eta}_y\left(\sigma_y \otimes\kappa_z\right)\right]+\sum_{A,B}\frac{\lambda_j}{2}\left[\left(\mathbbm{1}^A\pm\sigma_z\right)\otimes\left(\mathbbm{1}+\vec{e_r}.\vec{d}\right)\right]
\end{equation}
where $\eta$ is the band parameter, $\hat{\eta}_x$ and $\hat{\eta}_y$ are wave number operators and $\mathds{1}^{A(B)}$ is the $2 \times 2$ identity operator acting on the system $A(B)$. The intervalley and intravalley scattering processes are accommodated by parameters $\lambda_A$ and $\lambda_B$ along with $\vec{e_r}=(\cos\alpha,\sin\alpha,0)$ accompanied by a phase shift $\alpha$ and we consider the case $\lambda_A=\lambda_B=\lambda$. In the standard computational basis $\{ |\uparrow 1 \rangle,|\uparrow 0\rangle,|\downarrow 1 \rangle,|\downarrow 0\rangle\}$, the eigenvalues and the corresponding eigenvectors of the Hamiltonian are computed as, 
\begin{subequations}
\begin{align}
    \mathcal{E}_1=  \lambda\left(1-\sqrt{n_+^2+\eta_{22}^2}\right), ~& ~~~~~~\ket{\phi_1}= e^{-i\alpha}N_+\ket{00}+\frac{1}{2}\ket{01}+\frac{e^{-i\alpha}}{2}\ket{10}+N_+\ket{11},~\nonumber~~\\
   \mathcal{E}_2=  \lambda\left(1+\sqrt{n_+^2+\eta_{22}^2}\right), ~& ~~~~~~\ket{\phi_2}= -e^{-i\alpha}N_+\ket{00}+\frac{1}{2}\ket{01}+\frac{e^{-i\alpha}}{2}\ket{10}-N_+\ket{11},\nonumber \\
   \mathcal{E}_3=  \lambda\left(1-\sqrt{n_-^2+\eta_{22}^2}\right), ~& ~~~~~~\ket{\phi_3}= -\frac{e^{-i\alpha}}{2}\ket{00}+N_{-}\ket{01}-e^{-i\alpha}N_{-}\ket{10}+\frac{1}{2}\ket{11},\nonumber \\
   \mathcal{E}_4=  \lambda\left(1+\sqrt{n_-^2+\eta_{22}^2}\right), ~& ~~~~~~\ket{\phi_4}= -\frac{e^{-i\alpha}}{2}\ket{00}-N_{-}\ket{01}+e^{-i\alpha}N_{-}\ket{10}+\frac{1}{2}\ket{11},\nonumber
\end{align}
\end{subequations} 
where $n_{\pm}=\eta_{11}\pm 1$ and   the Dirac point $K(K')$is denoted by the state $\ket{ab}$ with $a=0(1)$ and for the sub-lattice A(B) with $b=0(1)$
with the coefficients $N_{\pm}$  being defined as 
\begin{equation}
    N_{\pm} = -\frac{n_{\pm} \mp i \eta_{22}}{2\sqrt{n_{\pm}^2 +\eta_{22}^2}},
\end{equation}
with $\eta_{11}=\frac{\eta\, \eta_x}{\lambda}$ \text{and} $\eta_{22}=\frac{\eta\, \eta_y}{\lambda}$. The concurrence, TMIN and UIN for the above states are unity implying that the eigenstates are maximally entangled (correlated).
\section{Quantum correlations in a graphene sheet}

In this section, we explore the behaviors of quantum correlations contained in a two-dimensional graphene sheet. The correlation in ground state of the graphene sheet is explored by the Byres measure of entanglement and thermal correlations measured in terms of Bures entanglement, trace MIN and UIN.

\subsection{Ground state properties}
\label{ground}

 The ground state of the system is decided based on the competing system parameters $\eta_{11}$ and $\lambda$. When $\lambda> 0$ and $\eta_{11} > 0 (\eta_{11} < 0)$, the ground state of the system is $|\phi_1\rangle (|\phi_3\rangle)$. Similarly, the ground of the system for the parametric space  $\lambda \leq 0$ and $\eta_{11}> 0 (\eta_{11}<0)$ is $|\phi_2\rangle (|\phi_4\rangle)$. Initially, the Bures entanglement is calculated for the energy eigenstates $\{\ket{\phi_i}\}$ of Eq. (10). The Bures entanglement is directly obtained from the concurrence \begin{equation*}
    C(\ket{\phi_1})=C(\ket{\phi_2})=\frac{|\eta_{22}|}{\sqrt{(1+\eta_{11})^2+ \eta^2_{22}}},
\end{equation*}
\begin{equation}
    C(\ket{\phi_3})=C(\ket{\phi_4})=\frac{|\eta_{22}|}{\sqrt{(1-\eta_{11})^2+ \eta^2_{22}}}.
\end{equation} 
 When $\eta_{11}=0$ and $\lambda>0$, the system is degenerate and the ground state is obtained as the linear combination of $\ket{\phi_1}$ and $\ket{\phi_3}$ as 
 \begin{equation}
   \ket{\phi_g^+}=\frac{1}{\sqrt{2}}(\hbox{e}^{\hbox{i}\beta_1}\ket{\phi_1}+\ket{\phi_3})
\end{equation} 
where $\beta_1$ is the relative phase in which the two states are equally likely  according to the principle of  maximum entropy in a microcanonical ensemble. On the other hand, when $\eta_{11}=0$ and $\lambda<0$, the ground state is a linear combination of $\ket{\phi_2}$ and $\ket{\phi_4}$, \begin{equation}
    \ket{\phi_g^-}=\frac{1}{\sqrt{2}}(\hbox{e}^{\hbox{i}\beta_2}\ket{\phi_2}+\ket{\phi_4}).
\end{equation}
with the relative phase factor being $\beta_2$.

 It is quite well-known that the entanglement of the superposition of maximally entangled states need not be maximal \cite{Linden2006}. For the relative phases $\beta_1=\beta_2=0$, the concurrence of the states $\ket{\phi_g^-}$ and $\ket{\phi_g^+}$ are equal and found to be $|\eta_{22}|/\sqrt{1+\eta_{22}^2}$. The correlation measures like concurrence, TMIN and UIN of the degenerate states $\ket{\phi_g^-}$ and $\ket{\phi_g^+}$ are equal and hence we consider the Bures entanglement measure alone for  further investigation on nonlocal properties of ground states.

\begin{figure*}[!ht]
\centering\includegraphics[width=0.46\linewidth]{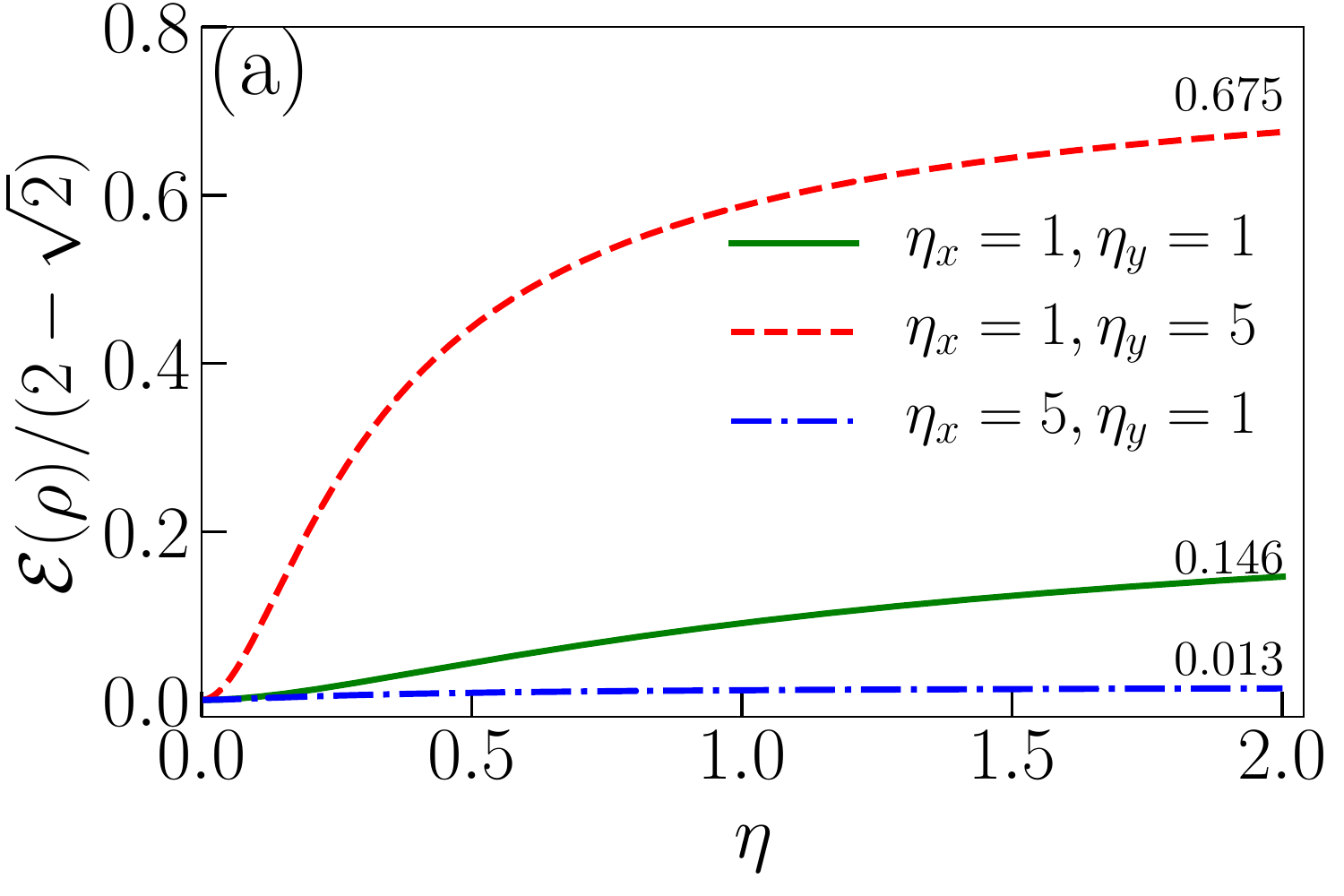}
\centering\includegraphics[width=0.46\linewidth]{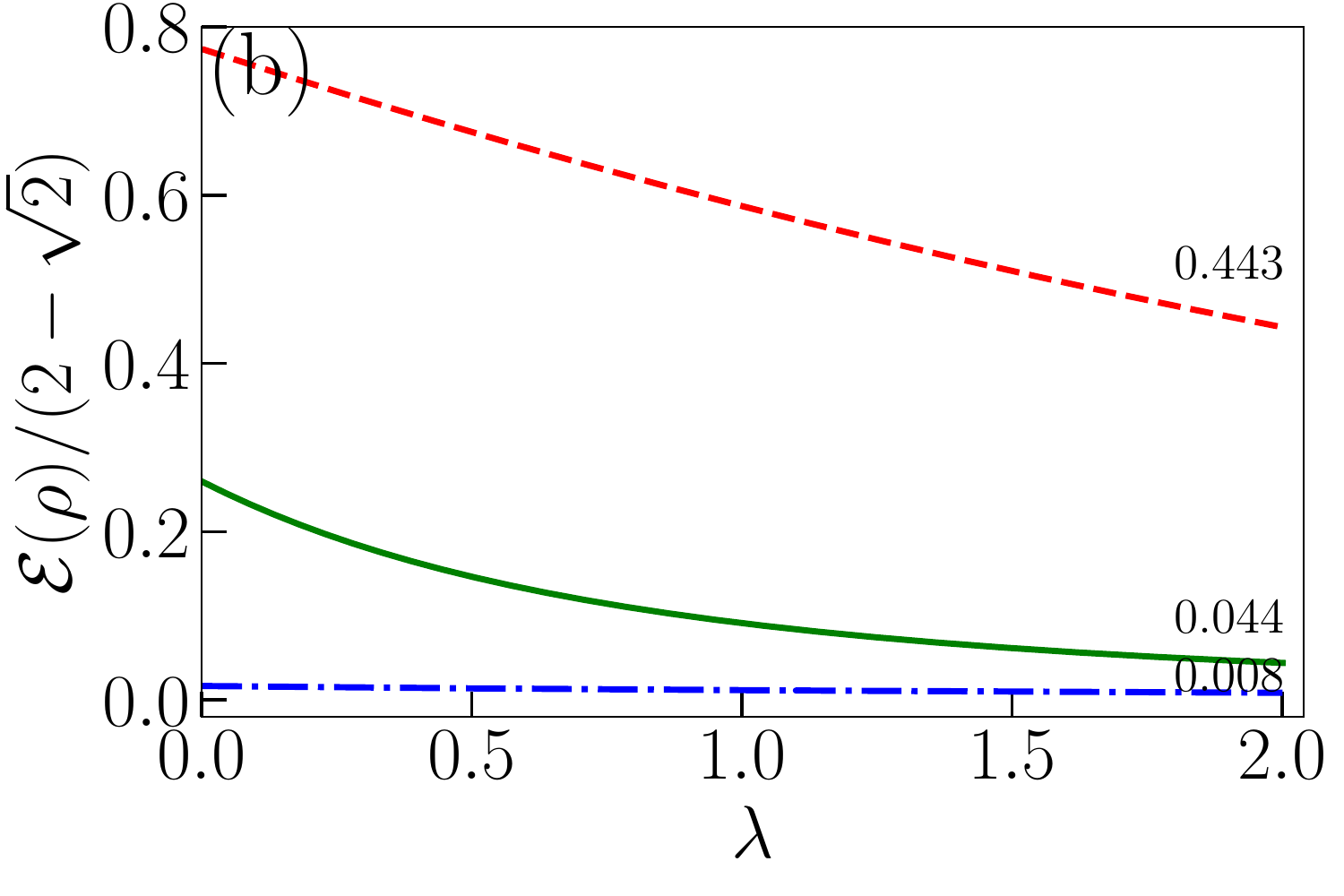}
\caption{(Color online) The ground state entanglement of graphene sheet quantified by Bures distance as a function of (a) band parameter ($\lambda=1$)  and (b) scattering strength ($\eta=1$).}
\label{fig2}
\end{figure*}

 In general, the Bures entanglement varies from $0$ (separable state/unentangled) to $2-\sqrt{2}$ (maximally entangled state). For the sake of easy comparison, we have normalised the Bures entanglement by dividing the factor $2-\sqrt{2}$. To understand the ground state entanglement properties, we have plotted the Bures entanglement of the state $\ket{\phi_1}$ or $\ket{\phi_2}$ in the graphene sheet as a function of band parameter $\eta$ and scattering strength  $\lambda$ in Fig. \ref{fig2}. It is quite obvious  that the degree of entanglement shows the monotonic behavior with the band parameter $\eta$ and scattering length $\lambda$. From Fig. \ref{fig2}, we observe that the entanglement is zero when the band parameter $\eta$ is zero, and entanglement increases with the increase of  $\eta$. It is evident that the entanglement was generated due to the increase of the band parameter. For higher values of $\eta$, the  entanglement between the lattice points reaches the maximum value. On the other hand, the entanglement is maximum at $\lambda=0$ and decreases with the increase of the scattering parameter (shown in Fig. \ref{fig2}b).

\begin{figure*}[!ht]
\centering\includegraphics[width=0.46\linewidth]{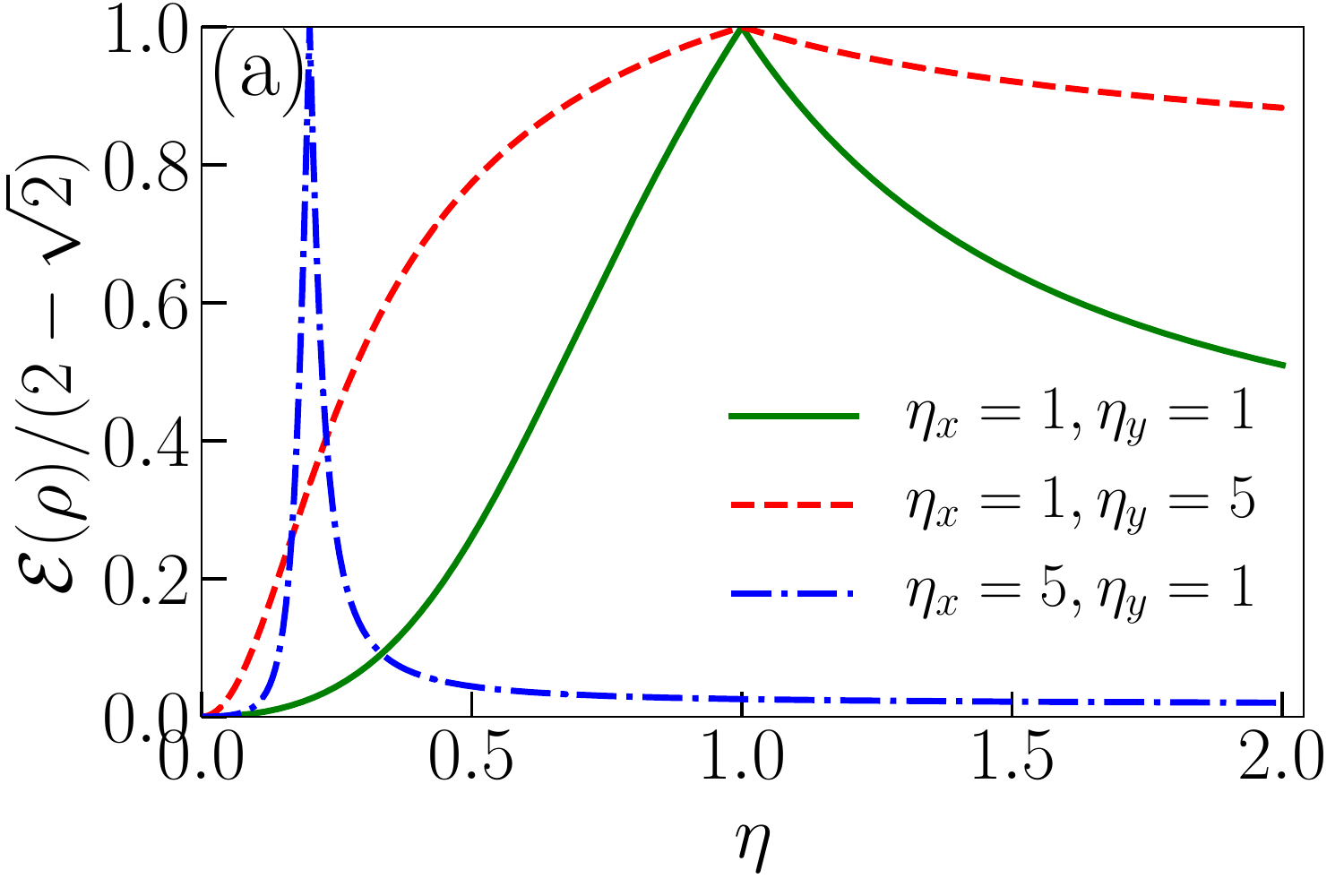}
\centering\includegraphics[width=0.46\linewidth]{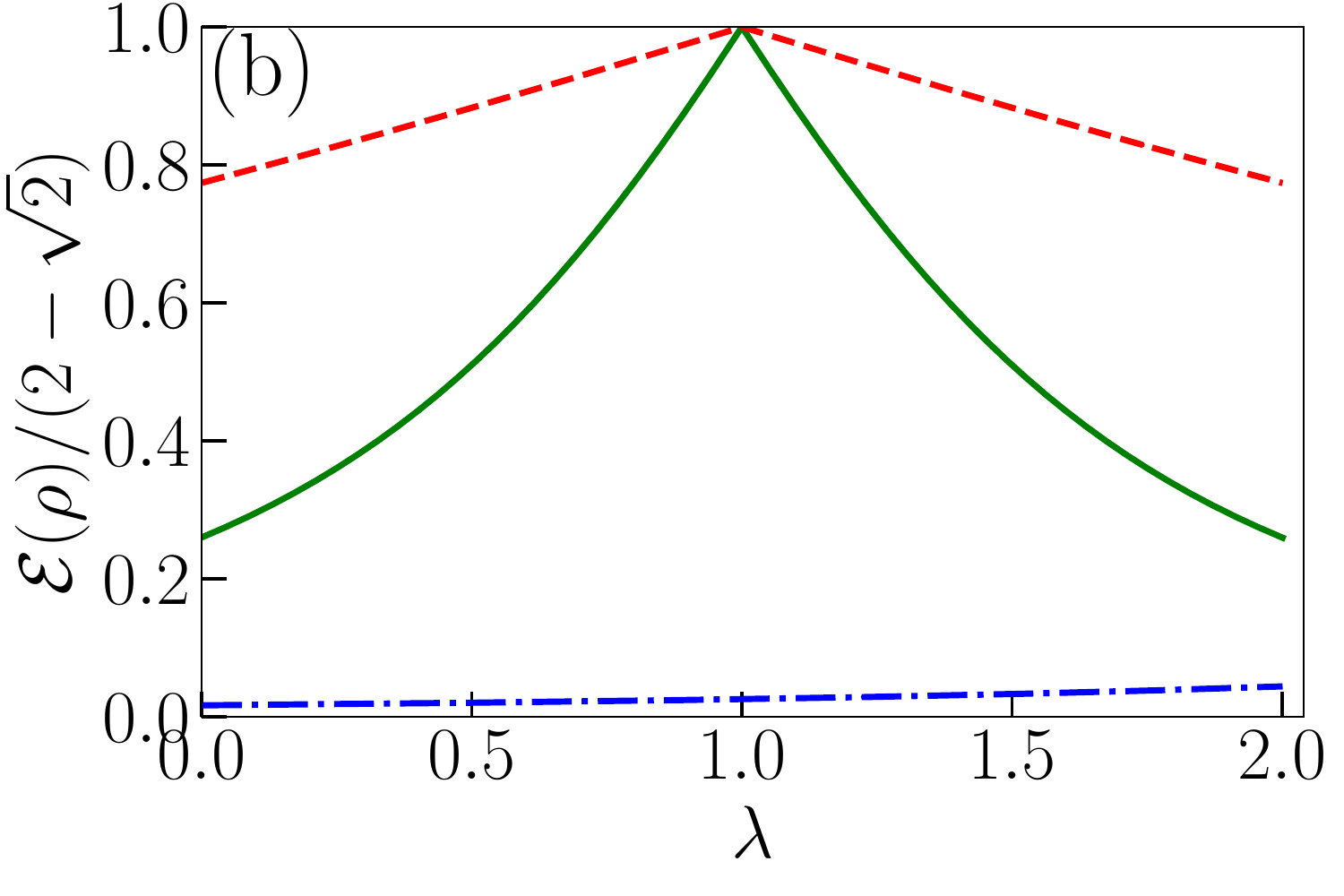}
\caption{(Color online) The ground state entanglement of graphene sheet quantified by Bures distance as a function of (a) band parameter ($\lambda=1$) and (b) scattering strength with $\eta=1$. }
\label{fig3}
\end{figure*}

Next, we study the behaviors of  Bures entanglement for another ground state $\ket{\phi_3}$ or $\ket{\phi_4}$ of the graphene system. The entanglement of the state $\ket{\phi_3}$ or $\ket{\phi_4}$ is plotted against $\eta$ and $\lambda$ in Fig. \ref{fig3}. Similar to the previous case, the entanglement is zero for $\eta=0$ and maximum when $\lambda$ is zero.   For constant values of  $\eta_y$, increase of  $\eta_x$  shows a decrease in Bures measure for the ground state whereas keeping  $\eta_x$, $\eta_y$ and $\lambda$ constant, it is observed that entanglement reduces for increasing $\eta$ after reaching a maximal entanglement (solid and dot-dashed curves in Fig. \ref{fig3}(a)).  The maximally entangled quantum states are obtained for states $\ket{\phi_4}$ and $\ket{\phi_3}$ setting the state parameters $\eta_x=\eta_y=\eta=\lambda=1$ (solid curve), $\eta_x=\eta=\lambda=1$, $\eta_y=5$ (dashed curve) and $\eta_y=\lambda=1$, $\eta_x=5$ and $\eta=0.2$ (dotted-dashed curve) as can be seen in Fig. \ref{fig3}(a). However, the maximal entanglement  can also be obtained by appropriately manipulating the system parameters as shown in Fig. \ref{fig3}b. This can be witnessed in the peak value of Bures measure between the region $0<\lambda<0.5$  for $\eta_x=5, \eta_y=1$ in Fig. \ref{fig3}. 


\subsection{Thermal quantum correlation}
\label{state}
To study the thermal effect on the selected quantum resources of the graphene sheet, we construct the thermal state of the system.  At thermal equilibrium, it is defined as 
\begin{align}
    \rho(T)=\frac{1}{Z}\exp[-\beta \hat{H}]
\end{align}
where $\beta=1/k_B T$ is the inverse temperature with $k_B$ being the Boltzmann constant and  $Z=\Tr\exp[-\beta \hat{H}]$ is the partition function. The thermal state of the system is obtained as
\begin{equation}
    \rho(T) = \frac{1}{Z}\left(\begin{array}{cccc}
\varrho_{11} & \varrho_{12} & \varrho_{13} & \varrho_{14} \\
\varrho^*_{12} & \varrho_{11} & \varrho_{23} & \varrho_{24} \\
\varrho^*_{13} & \varrho^*_{23} & \varrho_{11} & \varrho_{34} \\
\varrho^*_{14} & \varrho^*_{24} & \varrho^*_{34} & \varrho_{11}
\end{array}\right).
\label{thermal}
\end{equation}
The matrix elements are
\begin{subequations}\begin{eqnarray}
 \rho_{11} = \frac{Z}{4},~~\rho_{12}= \frac{1}{2}e^{-i\alpha}\left(e^{-\beta \mathcal{E}_1}N_+-Z' N_-^*\right), ~~\rho_{13}=\frac{1}{2}\left(e^{-\beta \mathcal{E}_1}N_++Z'N_-^*\right) \nonumber\\ 
 \rho_{14}=-\frac{1}{4}e^{-i\alpha}\left(-2e^{-\beta \mathcal{E}_1}+Z\right), \rho_{24}=\frac{1}{2}\left(e^{-\beta \mathcal{E}_1}N_+^*+Z' N_-\right), \nonumber \\
 \rho_{23}=-\frac{1}{4}e^{i\alpha}\left(-2e^{-\beta \mathcal{E}_1}+Z\right), \rho_{34}= \frac{1}{2}e^{-i\alpha}\left(e^{-\beta \mathcal{E}_1}N_+^*-Z' N_-\right) \nonumber
 \end{eqnarray}
 \end{subequations}
 with $Z = \sum_ie^{-\beta \mathcal{E}_i}$ and $Z'=e^{-\beta \mathcal{E}_2}+e^{-\beta \mathcal{E}_3}-e^{-\beta \mathcal{E}_4}.$

\begin{figure*}[!ht]
\centering\includegraphics[width=0.4\linewidth]{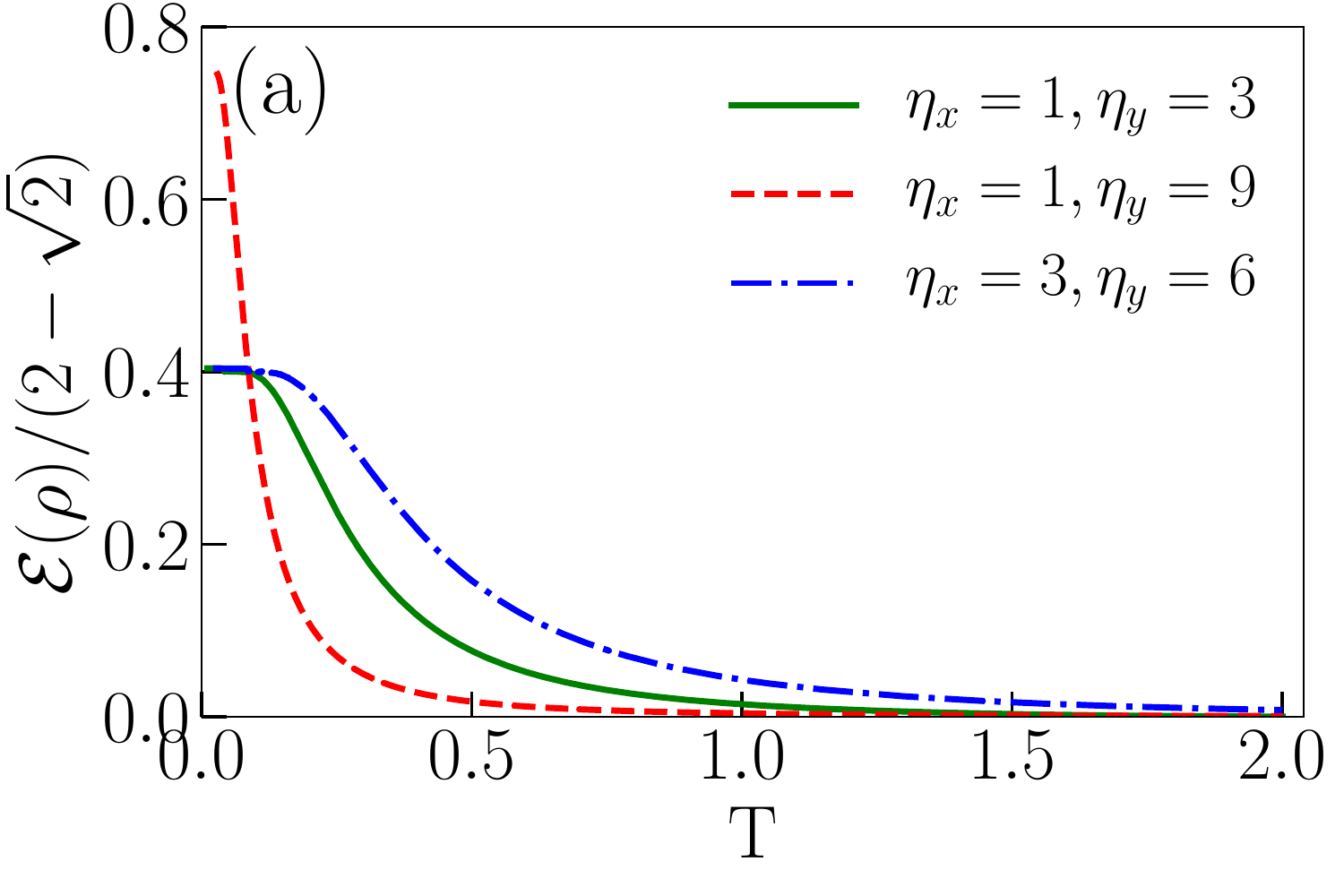}
\centering\includegraphics[width=0.4\linewidth]{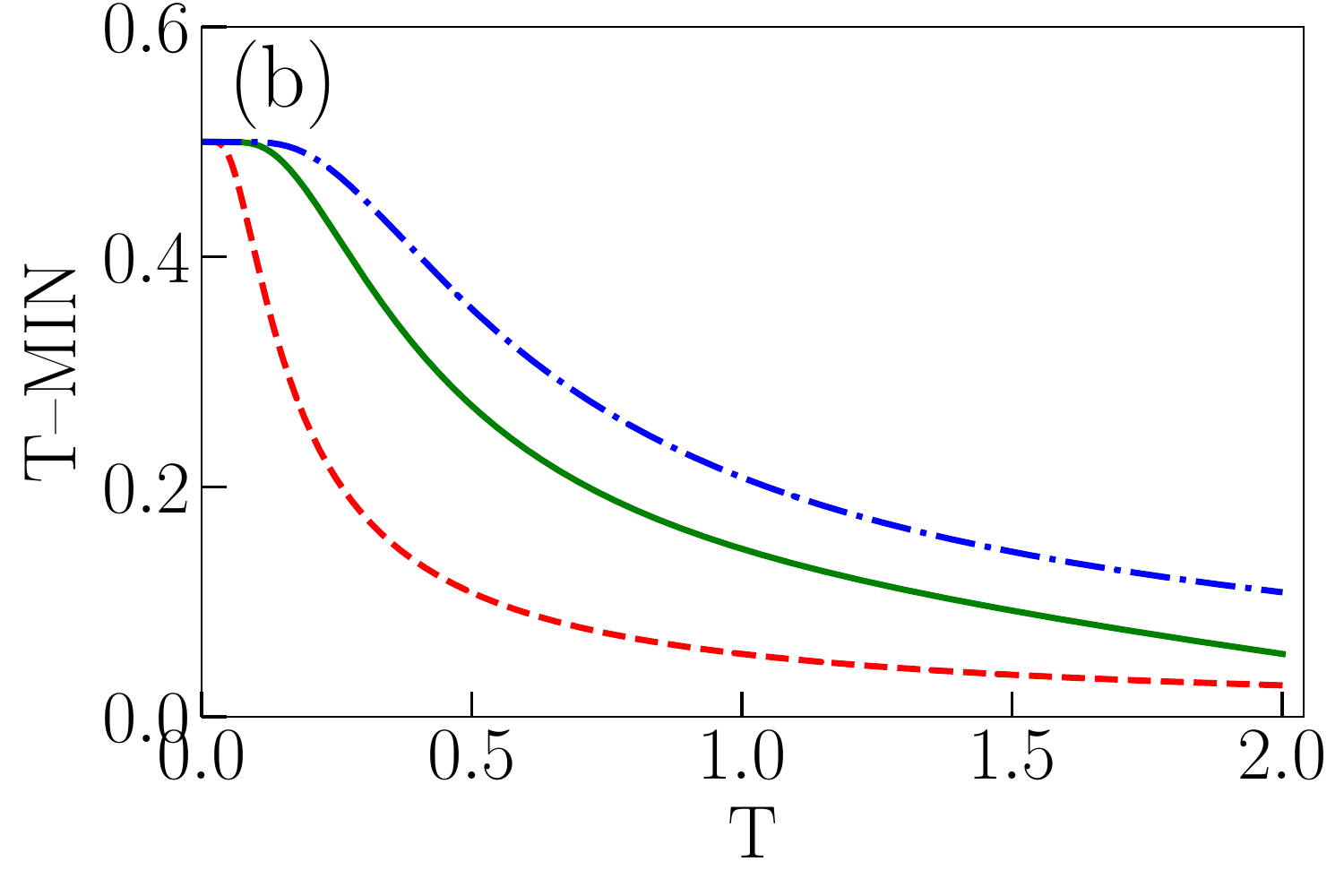}
\centering\includegraphics[width=0.4\linewidth]{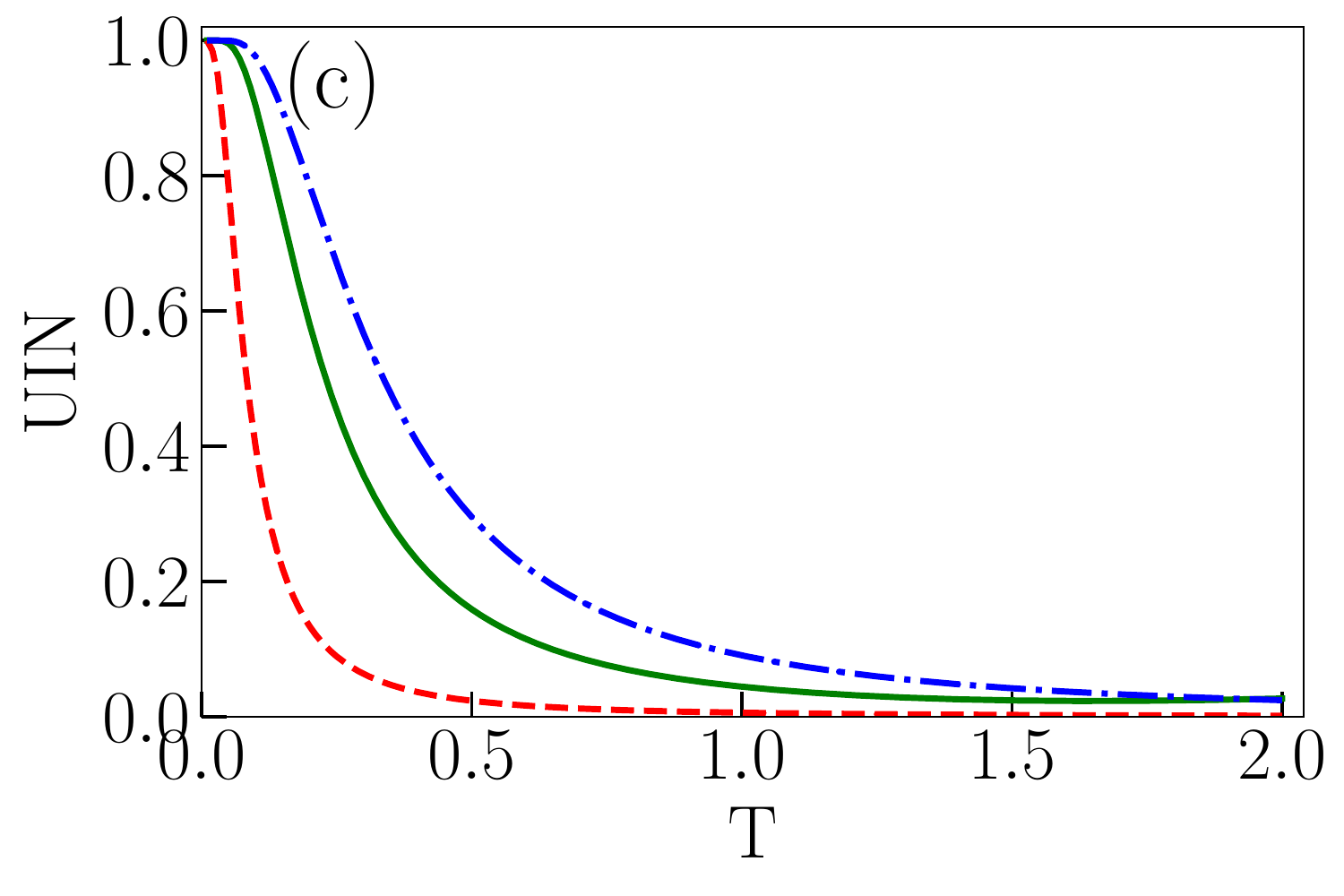}
\caption{(color online) The thermal quantum correlations of two-dimensional graphene sheet as a function of temperature (a) Bures Entanglement (b) Trace distance MIN, and  (c) UIN for  the  parametric choice  $\eta=1$, $\lambda=1$, and $\alpha=\pi/3$. }
\label{fig4}
\end{figure*}

 The analytical expressions of the quantum correlation measures are complicated and hence we resort to qualitative discussions. Here, we study the effects of band parameters and scattering strength on thermal quantum correlations quantified by Bures entanglement,  trace MIN (TMIN) and UIN. In Fig. \ref{fig4}, we have plotted the behavior of quantum correlations as a function of temperature $T$ for a fixed   band parameter and scattering strength. At $T=0$, the correlations between the lattice points are maximum  and decrease with the temperature $T$. A similar observation is observed for entanglement in terms of concurrence in Ref. \cite{HU2009}.  As we increase the temperature, all the correlation measures  decrease from the maximal value  and almost vanish at  sufficiently higher temperatures (critical temperature $T_c$). While we increase the value of  $\eta$, the Bures entanglement is sustained at higher temperatures when compared to the other associated quantities plotted in Fig. \ref{fig4}(b) \& (c).  In other words, the Bures entanglement offers more resistance to thermal fluctuations than trace MIN and UIN.  Increasing the wave number $\eta_x$ reduces the decay of correlations (dot-dashed curve) while higher $\eta_y$ causes the correlations to decrease rapidly (dashed curve). 

\begin{figure*}[!ht]
\centering\includegraphics[width=0.46\linewidth]{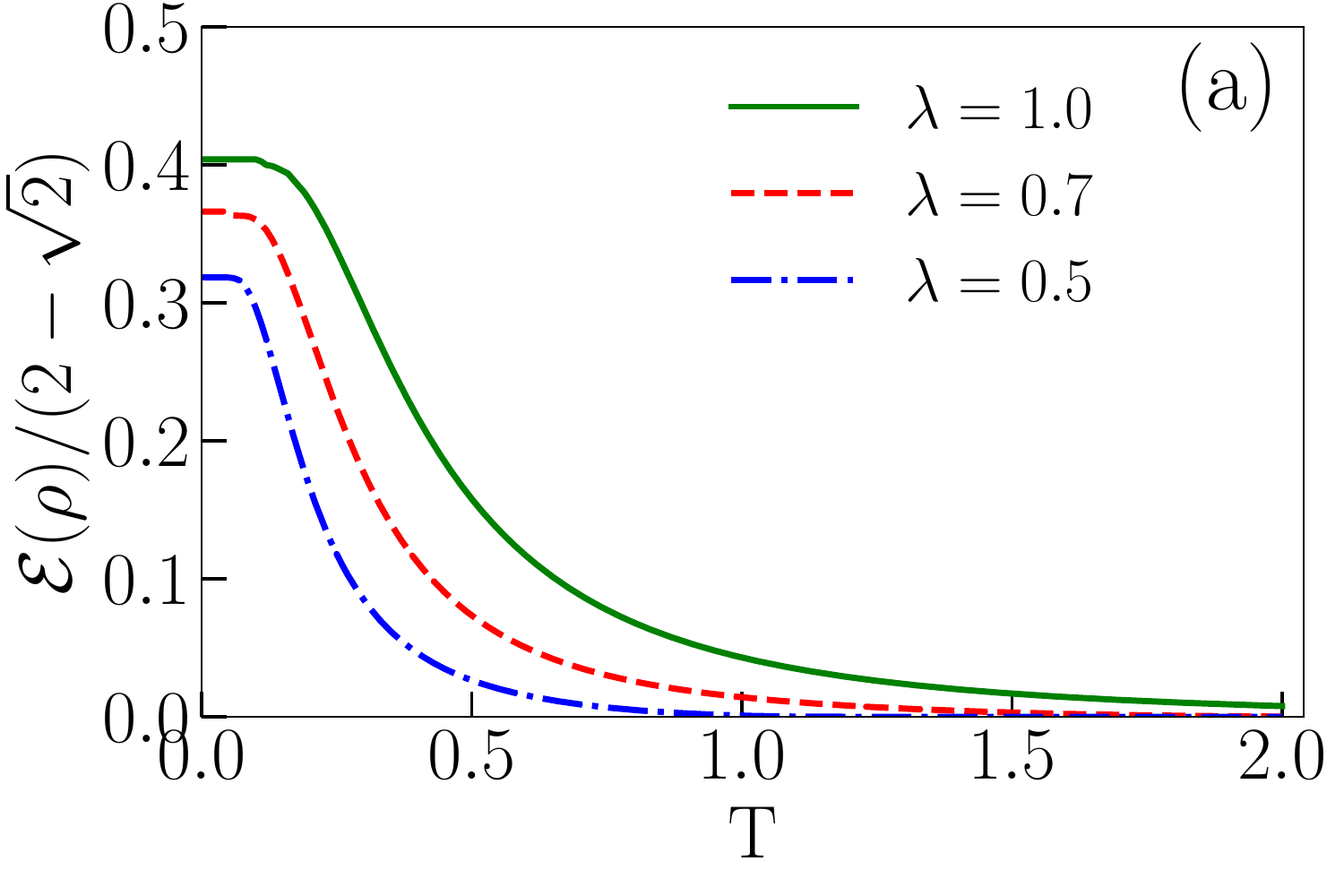}
\centering\includegraphics[width=0.46\linewidth]{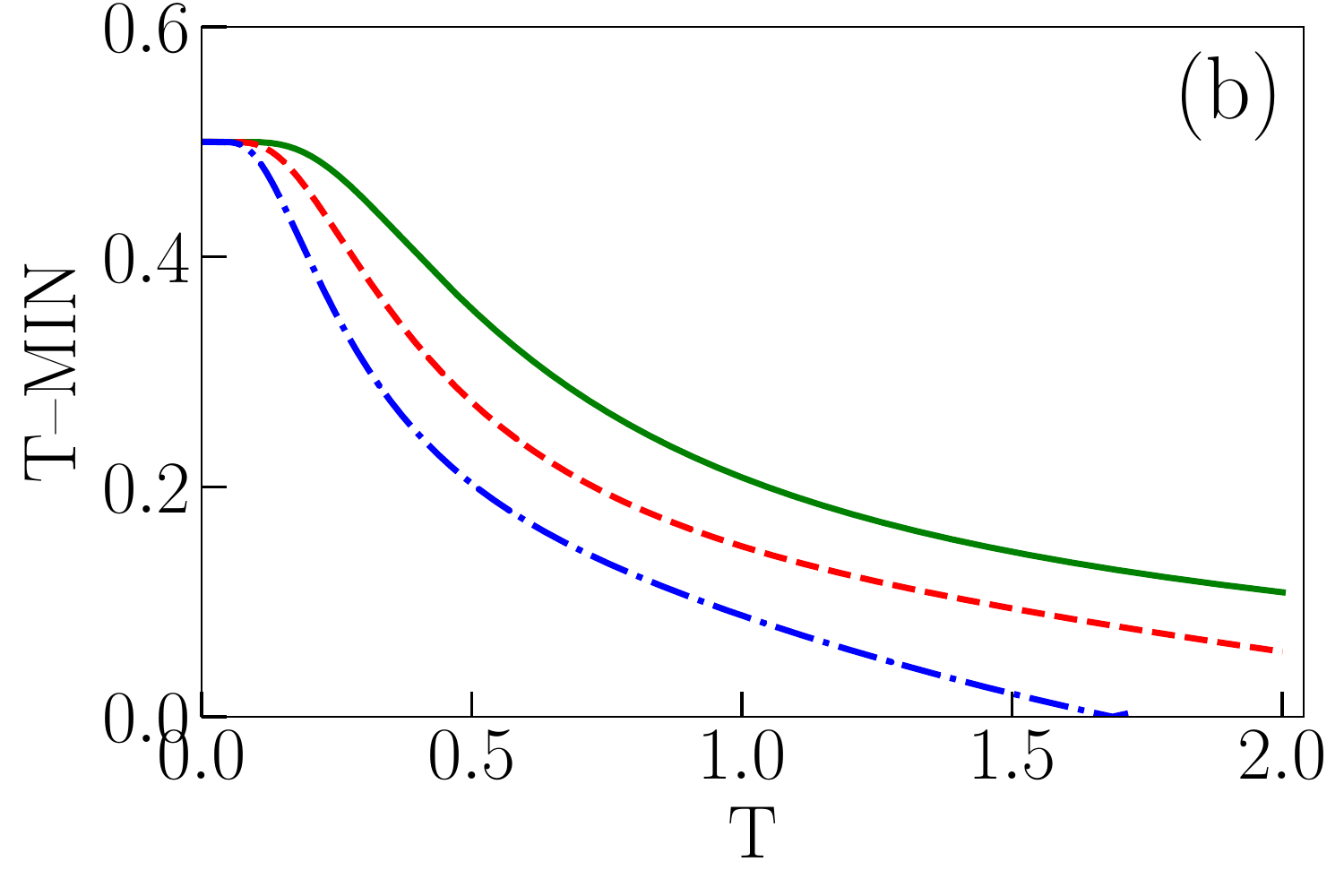}
\centering\includegraphics[width=0.46\linewidth]{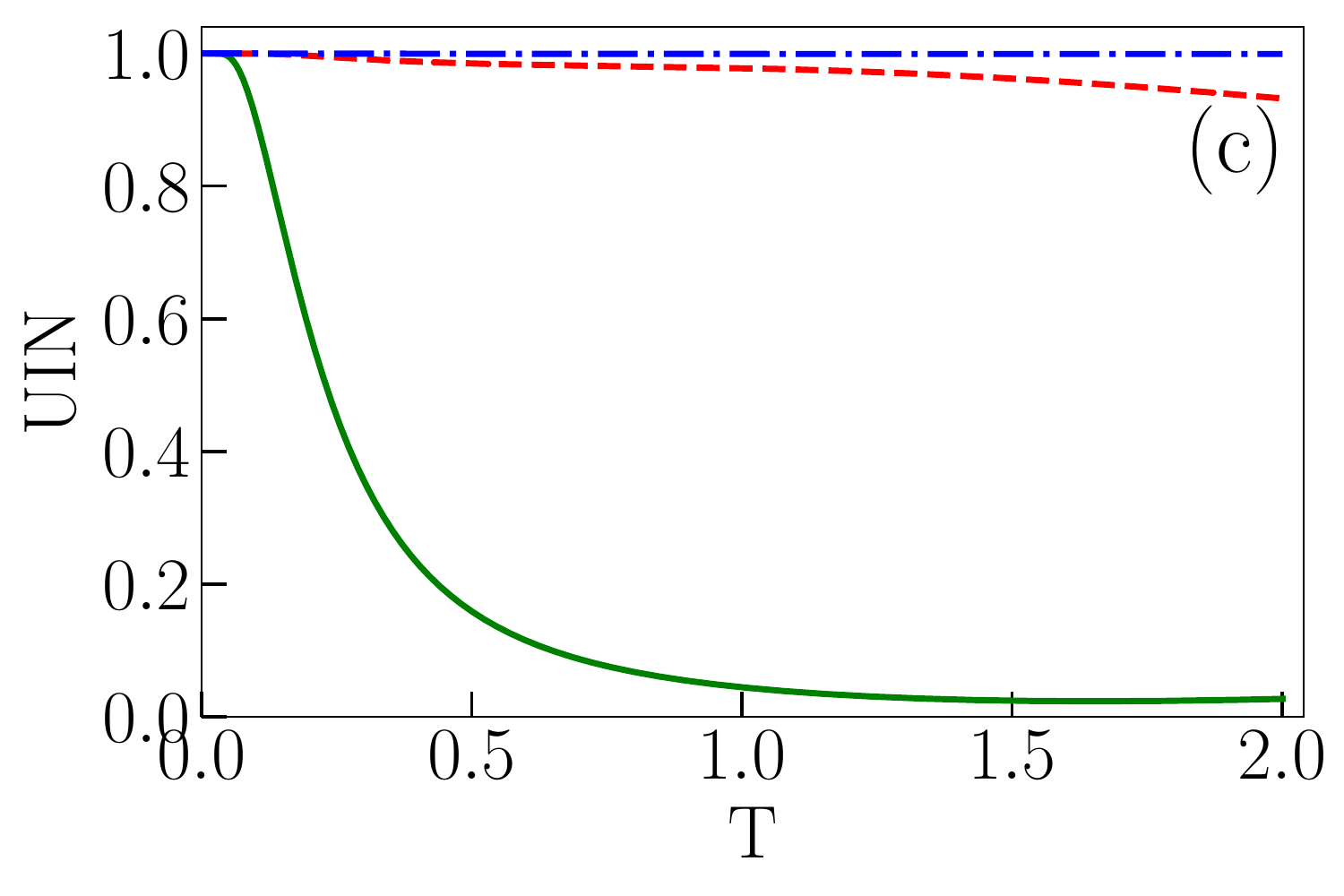}
\caption{(color online) The variation of quantum correlations with temperature for different values of scattering coefficients (a) Bures entanglement, (b) Trace distance MIN,  and (c) UIN for the parametric choice  $\eta=1$, $\alpha=\pi/3$, $\eta_x=3, \text{and}~ \eta_y=6$. }
\label{fig5}
\end{figure*}

Then, the role of scattering parameter $\lambda$ is studied on the correlation quantifiers in Fig. \ref{fig5}.  All the quantum correlation measures monotonically decrease with the increase of temperature. In general, we observe that the  critical temperature where correlations tend to zero can be increased by decreasing $\lambda$. The decay of UIN is more sensitive to $\lambda$ compared to Bures measure of entanglement and TMIN. 

\section{Teleportation}
\label{Telepor}
 Teleportation is one of the most well-known quantum information processing tasks that exploit the entanglement as a resource and its goal is to teleport an arbitrary quantum state. The story of quantum teleportation is that party $a$ wishes to send an arbitrary quantum state to party $b$. They agree to use a quantum channel by a pair of entangled states and share it between themselves.  Here, we consider the thermal state $\rho(T)$ of the graphene sheet as a channel for teleportation. For an arbitrary two-qubit pure state, the input unknown state can be considered as
\begin{equation}
    \ket{\Psi_{in}}=\hbox{cos}\left(\frac{\theta}{2}\right) \ket{10}+\hbox{sin}\left(\frac{\theta}{2}\right)\ket{01},\ \ (0\leq\theta\leq\pi),
\end{equation}
where $\theta$ is the amplitude  of the state. Then, the output state is defined as 
\begin{align}
    \rho_{out}=\sum_{i,j=0,x,y,z}\mathcal{P}_{ij}\left(\sigma_i \otimes \sigma_j\right) \rho_{in} \left(\sigma_i \otimes \sigma_j\right),
    \label{channel}
\end{align}
where $\rho_{in}=\ket{\Psi_{in}}\bra{\Psi_{in}}$, $\sigma_0=\mathds{1}$ and $\sigma_j (j=x,y,z)$ are the usual Pauli's spin matrices. Further, 
\begin{equation}
   \mathcal{P}_{ij}=\mathcal{P}_{i}\mathcal{P}_{j}=\Tr{E^i\rho_{ch}} \Tr{E^j\rho_{ch}}
\end{equation} 
where $\sum_{i,j}\mathcal{P}_{ij}=1$. Herein, $E^0=\ket{\Psi^-}\bra{\Psi^-},\,E^1=\ket{\Phi^-}\bra{\Phi^-},\,E^2=\ket{\Phi^+}\bra{\Phi^+},\,E^3=\ket{\Psi^+}\bra{\Psi^+}$
with  $\ket{\Psi^{\pm}}$ and $\ket{\Phi^{\pm}}$ being the Bell channels. Using  Eqs. (\ref{channel}) and (\ref{thermal}), we obtain the output density matrix in the form
\begin{align}
    \rho_{out} = \left(\begin{array}{cccc}
a & 0 & 0 &  b ~\text{sin}\theta \\
0 & a &  b ~\text{sin}\theta & 0 \\
0 &  b ~\text{sin}\theta & a & 0 \\
 b ~\text{sin}\theta & 0 & 0 &  a
\end{array}\right)
\end{align}
where the matrix elements are 
\begin{align}
  a = \frac{1}{4}, ~~~
 \text{and}~~~
 b =\frac{ \cos ^2(\alpha ) \left(\cosh \left(\frac{\lambda  \sqrt{n_-^2+n_{22}^2}}{T}\right)-\sinh \left(\frac{\lambda  \sqrt{n_+^2+n_{22}^2}}{T}\right)\right)^2}{4 \left(\cosh \left(\frac{\lambda  \sqrt{n_-^2+n_{22}^2}}{T}\right)+\cosh \left(\frac{\lambda  \sqrt{n_+^2+\text{n22}^2}}{T}\right)\right)^2} \nonumber
\end{align}
with $n_{\pm}=\eta_{11}\pm1$.
The efficiency of the teleportation technique is characterized by the fidelity measure $ \mathcal{F}(\rho_{out},\rho_{in})$ \cite{Jozsa} defined as 
\begin{align}
    \mathcal{F}(\rho_{out},\rho_{in})=\left(\Tr{\sqrt{(\rho_{in})^{1/2} \rho_{out} (\rho_{in})^{1/2}}}\right)^2. 
\end{align}
Further, this measure can be generalized for pure input states by defining average fidelity as \cite{Bowdrey2002}
\begin{align}
    \mathcal{F}_A = \frac{1}{4\pi} \int_0^{2\pi}\hbox{d}\phi\int_0^{\pi} \mathcal{F} \sin(\theta) \hbox{d}\theta. \nonumber
\end{align}
It is worth pointing out  that the classical average fidelity bounds occur at $ \mathcal{F}_A=2/3$ and if   $ \mathcal{F}_A>2/3$, then, the teleportation is achieved successfully. 

\begin{figure*}[!ht]
\centering\includegraphics[width=0.46\linewidth]{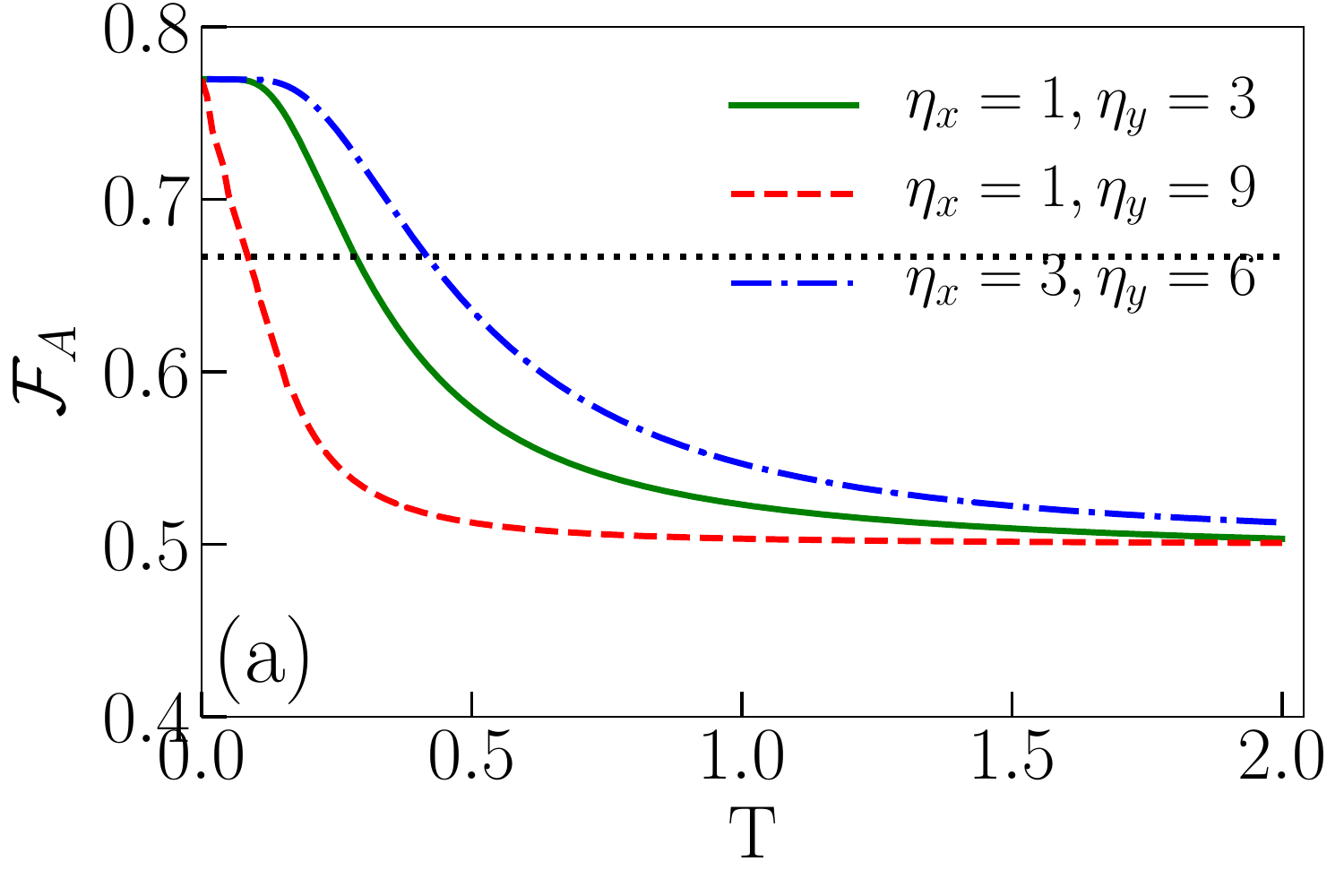}
\centering\includegraphics[width=0.46\linewidth]{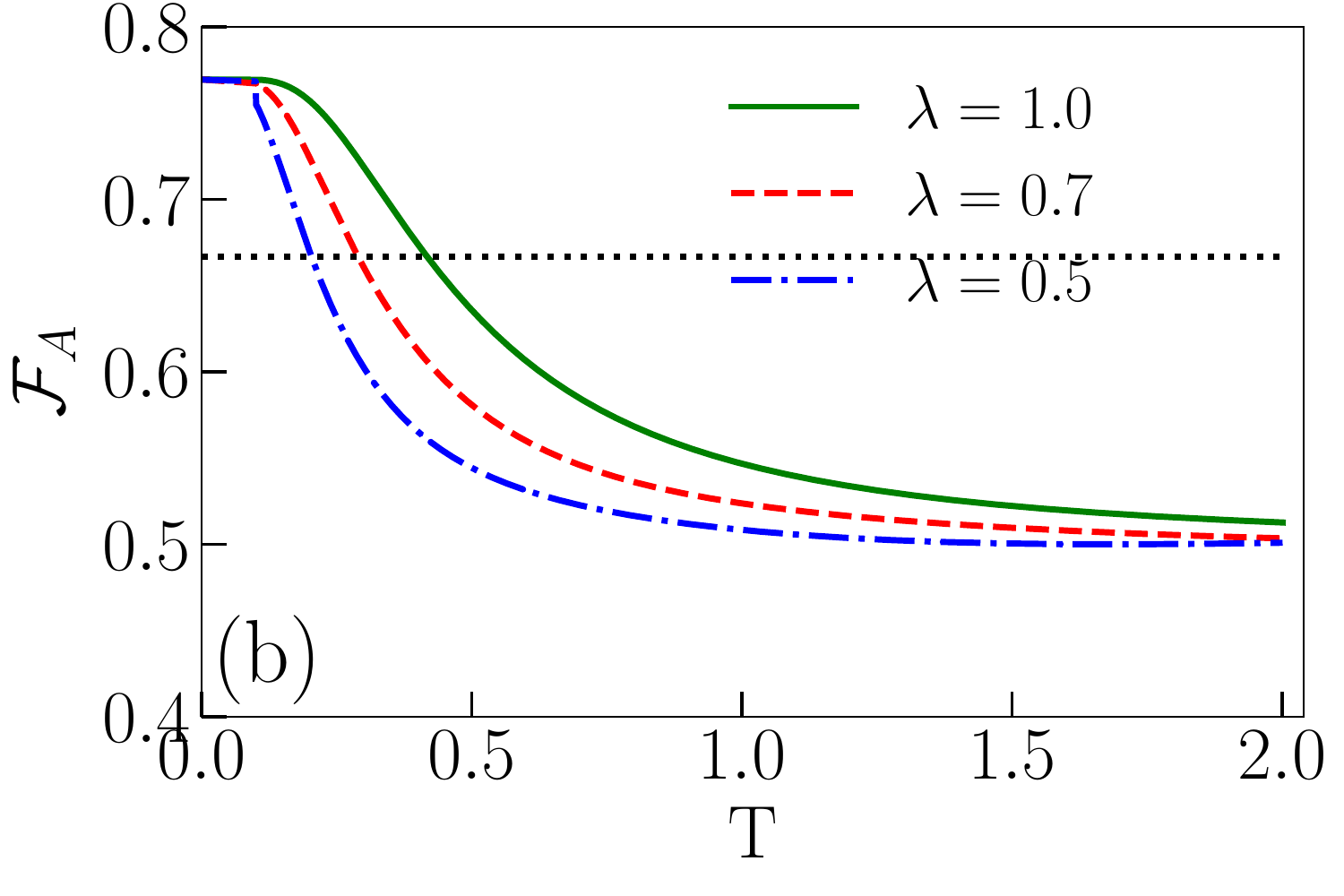}
\caption{(color online) Average fidelity of quantum teleportation of the quantum state as a function of temperature $T$ for  different values of (a) wave numbers and (b) scattering strength. The fixed parameters are $\eta=1$, $\alpha=\pi$,  and  $\theta=\pi/2$.  The dotted line separates the classical average fidelity and quantum fidelity domain $(\mathcal{F}_A>2/3)$.}
\label{fig6}
\end{figure*}

Now, we list below the main results obtained for the teleportation protocol by focusing on the role of band parameter and scattering coefficient. In Fig. \ref{fig6}, we discuss the average fidelity  as a function of the band parameter against temperature for a system of two qubits initially prepared in a pure state. We observe that the average fidelity decreases monotonically with the increase of temperature $T$. For $\eta_x=1, \eta_y=3 ~~\text{and}~ \lambda=1$, the maximum average fidelity can be obtained at $ \mathcal{F}_A>0.67$ for $T\rightarrow0$ and  is independent of the concurrence $C_{in}$ of the input state. It can be  noticed that quantum teleportation fails upon increasing the temperature and its failure is more pronounced   with the  increase of  wave number $\eta_y$. Hence,  manipulation of  appropriate parameters and the thermal state of the graphene sheet can be employed for teleportation. 

\begin{figure*}[!ht]
\centering\includegraphics[width=0.3\linewidth]{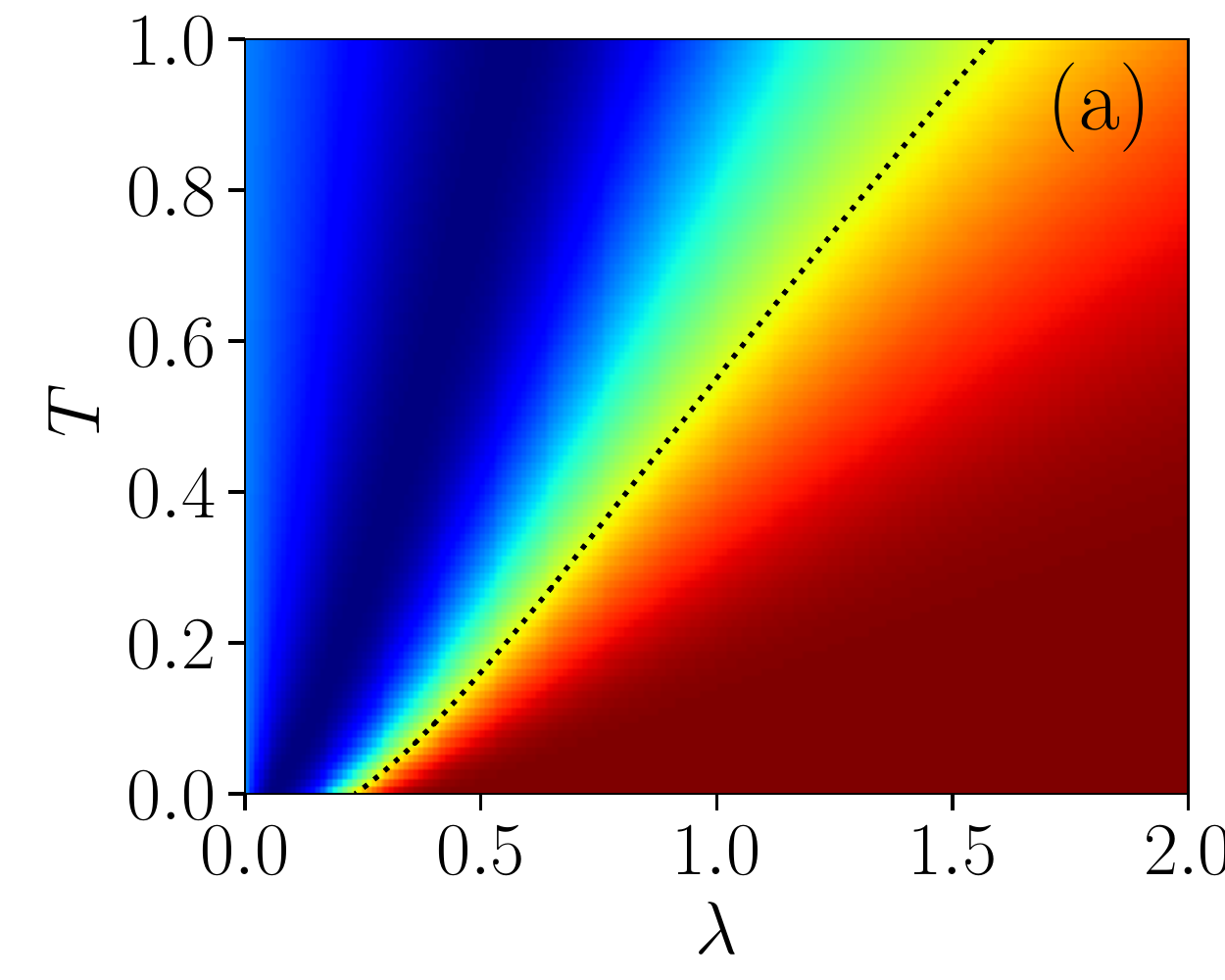}
\centering\includegraphics[width=0.3\linewidth]{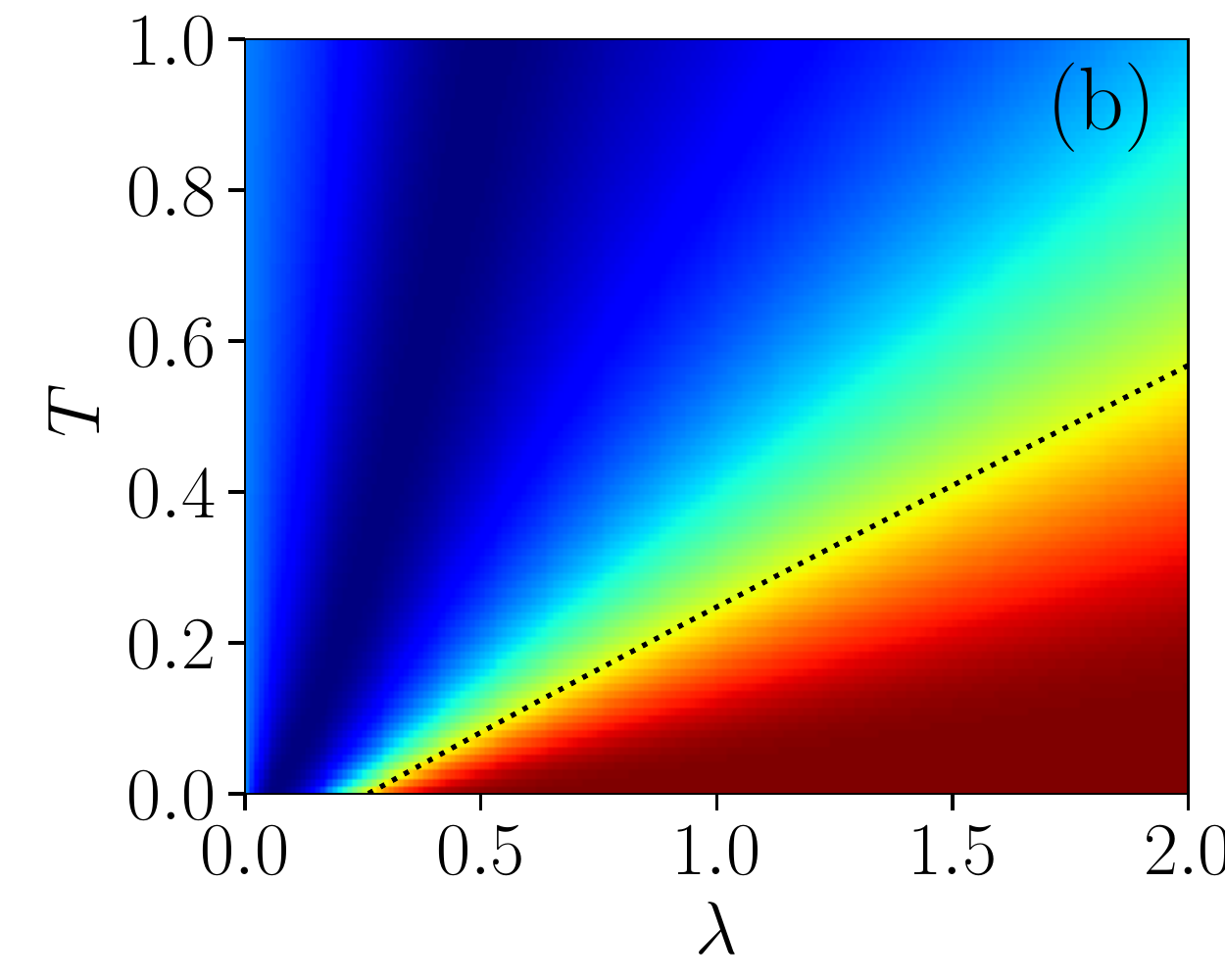}
\centering\includegraphics[width=0.3\linewidth]{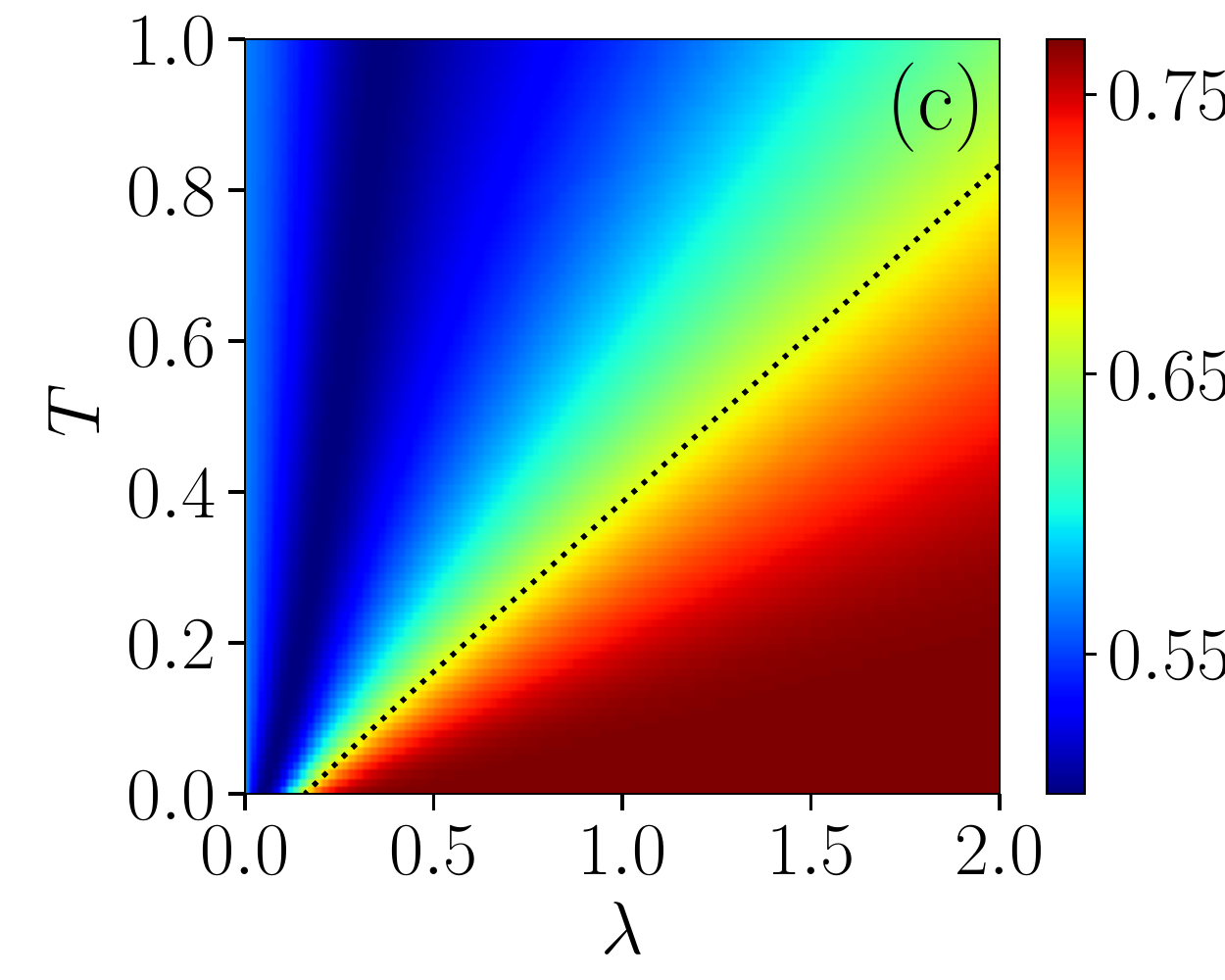}
\caption{(Color online) The density of average fidelity as a function of temperature and scattering strength for  (a) $\eta_x=1, \eta_y=1$, (b) $\eta_x=3, \eta_y=1$, (c) $\eta_x=1, \eta_y=3$. The fixed parameters are  $\eta=1$,  $\alpha=\pi$, and $\theta=\pi/2$.  The dotted line separates the classical average fidelity and quantum fidelity domain $(\mathcal{F}_A>2/3)$.}
\label{fig7}
\end{figure*}

Next, we study the role of  scattering strength on the performance of the teleportation process. In Fig. \ref{fig7}, we have plotted the density of average fidelity as functions of temperature $T$ and scattering strength $\lambda$. Here also, for a fixed  band parameter, we observe that the average fidelity is a decreasing function of the temperature. In other words, we find that at  higher values of  scattering strength  $\lambda$ and lower values of wave number $\eta_y$, the graphene sheet  remains a versatile resource to achieve successful teleportation of quantum state. In other words, higher scattering strength and the lowered wave number $\eta_y$ favors  average fidelity.

\section{Conclusion}
\label{cncl}
To summarize, we have investigated the quantum correlations measured by Bures entanglement, measurement-induced nonlocality (MIN) and uncertainty-induced nonlocality (UIN) between honeycomb lattice and Dirac points in a two-dimensional graphene sheet. In the ground state, we have evaluated all the correlation quantifiers analytically and the  concurrence, trace MIN and UIN are found to be equal.  The effects of the band parameter, the intravalley scattering process and the wave number on the ground state properties of the graphene are studied.

In addition, at thermal equilibrium,  we have also explored the thermal quantum correlations of the lattice points. It is shown that the band parameter strengthens the quantum correlations whereas the scattering strength weakens the correlation. Further, we have studied the teleportation of an unknown quantum state using the graphene sheet. The effects of system parameters on the teleportation are also brought out at thermal equilibrium.


\section*{Acknowledgment}
RM is grateful for the CTU Global Postdoc program and the financial support from MŠMT RVO 14 000. RR wishes to thank  the Council of Scientific and Industrial Research (CSIR), Government of India for financial support under Grant No. 03(1456)/19/EMR-II.

\end{document}